\documentstyle[editedvolume,numreferences]{crckapb} 

\input psfig.tex

\newcommand{\la}{\mathrel{\hbox{\rlap{\hbox{\lower4pt\hbox{$\sim$}}}\hbox{$<$}}}}
\newcommand{\ga}{\mathrel{\hbox{\rlap{\hbox{\lower4pt\hbox{$\sim$}}}\hbox{$>$}}}}

\begin{opening}
\title{Towards a Better Understanding of \protect \\ 
Active Galactic Nuclei$^1$}

\author{PAOLO PADOVANI}
\institute{Dipartimento di Fisica, II Universit\`a di Roma ``Tor Vergata''\\
           Via della Ricerca Scientifica 1, I-00133 Roma, Italy\\
           E-mail: padovani@roma2.infn.it}

\end{opening}

\runningtitle{TOWARDS A BETTER UNDERSTANDING OF AGN}

\begin{document}

\addtocounter{footnote}{1}
\footnotetext{Invited Review, to appear in {\it New Horizons from 
Multi-Wavelength Sky Surveys}, IAU Symposium 179, McLean B. et al. (eds.)}

\begin{abstract}
Active Galactic Nuclei (AGN) are ideal sources for multi-wave\-length
studies as their emission can cover almost 20 orders of magnitude in
frequency from the radio to the $\gamma$-ray band. After reviewing their basic
properties, I will assess how well we know the multifrequency spectra of AGN
as a class. I will then briefly illustrate how currently available and 
forthcoming sky surveys will help in addressing some of the open questions of
AGN studies. Finally, an analysis of the problem of the missing Type 2 QSO
will exemplify the dangers of monochromatic sky surveys for AGN. 
\end{abstract}

\section{AGN and Unified Schemes}
The large number of classes and subclasses which appear in AGN literature
might disorientate astronomers working in other fields. A simplified
classification, however, can be made based on radio loudness and the width of
the emission lines (e.g., \cite{UP95}). Radio-loud (RL) sources, that is
objects with radio to optical ($B$-band) flux ratio $f_{\rm r}/f_{\rm B} \ga
10$ (e.g., \cite{ST92}), make up only $\sim 10 - 15\%$ of AGN and have, for
the same optical luminosity, radio powers 3 to 4 orders of magnitude larger
than those typical of radio-quiet (RQ) sources. AGN are also divided in Type 1
(broad-lined) and Type 2 (narrow-lined) objects according to their
line-widths, with 1000 km/s (full width half maximum) being the dividing
value. We then have RL Type 1 AGN, that is radio quasars and broad-line radio
galaxies, and RL Type 2 AGN, that is radio galaxies, with their corresponding
RQ counterparts, that is Seyfert 1 galaxies and QSO, and Seyfert 2 galaxies
respectively. Some objects exist with unusual emission line properties, such as
BL Lacs, which have very weak emission lines with typical equivalent
widths $< 5$ \AA. 

In recent years we have come to understand that classes of apparently
different AGN might actually be intrinsically similar, only seen at different
angles with respect to the line of sight. The basic idea, based on a variety
of observations and summarized in Figure 1 of \cite{UP95}, is that emission in
the inner parts of AGN is highly anisotropic. The current paradigm for AGN
includes a central engine, surrounded by an accretion disk and by fast-moving 
clouds, probably under the influence of the strong gravitational
field, emitting Doppler-broadened lines. More distant clouds emit narrower
lines. Absorbing material in some flattened configuration (usually idealized
as a torus) obscures the central parts, so that for transverse lines of sight
only the narrow-line emitting clouds are seen (Type 2 AGN), whereas the near-IR
to soft-X-ray nuclear continuum and broad-lines are visible only when viewed
face-on (Type 1 AGN). In RL objects we have the additional presence of a
relativistic jet, roughly perpendicular to the disk, which produces strong
anisotropy and amplification of the continuum emission (``relativistic
beaming''). In general, different components are dominant at different
wavelengths. Namely, the jet dominates at radio and $\gamma$-ray frequencies
(although it does contribute to the emission in other bands as well), the
accretion disk is thought to be a strong optical/UV/soft X-ray emitter, while
the absorbing material will emit predominantly in the IR. 

This axisymmetric model of AGN implies widely different observational
properties (and therefore classifications) at different aspect angles. Hence
the need for ``Unified Schemes'' which look at intrinsic, isotropic properties,
to unify fundamentally identical (but apparently different) classes of AGN. 
Seyfert 2 galaxies have been ``unified'' with Seyfert 1 galaxies, whilst
low-luminosity and high-luminosity radio galaxies have been unified with BL
Lacs and radio quasars respectively (see \cite{A93} and \cite{UP95} and
references therein). 

\section{The multiwavelength spectrum of AGN}
The property that makes AGN ideal sources for multiwavelength studies is their
broad-band emission, which covers basically the whole observable electromagnetic
spectrum from the radio to the $\gamma$-ray band (almost 20 orders of magnitude
in frequency). Multifrequency coverage at this level, however, is rare 
for a single object: see for example \cite{A95} and \cite{L95} which give
simultaneous multifrequency data for NGC 3783, a Seyfert 1 galaxy, and 3C 273,
a radio-loud quasar, respectively. Also, the objects studied are not
necessarily representative of their class and are usually relatively local. 

Single-object studies are certainly important, as specific models can be
fitted to their multifrequency spectra to constrain the emission processes.
However, one would also like to learn about the general properties of the AGN 
population and study statistically their emission in various bands, for 
example to constrain unified schemes. In other words, use multifrequency data
to address some of the open questions of AGN research. 

Before I address this point, we first have to assess how well we know the
multifrequency spectrum of AGN as a class. I have then taken the latest
V\'eron-Cetty \& V\'eron AGN catalog (\cite{VV96}), which includes 11,662 AGN
and gives redshift, optical (U, B, and V) magnitudes, and some radio information
(6 and 11 cm fluxes), and cross-correlated it with radio, far/mid-IR, X-ray and
$\gamma$-ray catalogs available in machine readable form (using BROWSE as
implemented at SAX/SDC). The list includes the FIRST and NVSS (respectively
about 15\% and 40\% completed), the NORTH 20cm, PKS, PMN, GB6, S4, S5, 1 Jy
catalogs (radio: 20, 11, and 6 cm), the IRAS PSC and FSC (infrared: $12 - 100
\mu$), the {\it EXOSAT} CMA, the {\it ROSAT} RASS-BSC and WGA, the {\it
Einstein} Slew, EMSS, IPC catalogs (soft X-ray: $\sim 0.05 - 3.5$ keV), a hard
X-ray compilation (\cite{MB96}), and the {\it GRO} EGRET catalog ($\gamma$-ray:
30 MeV -- 20 GeV). Many of these surveys have been discussed at this meeting by
various contributions, where detailed references can be found. 

\begin{figure}
\psfig{file=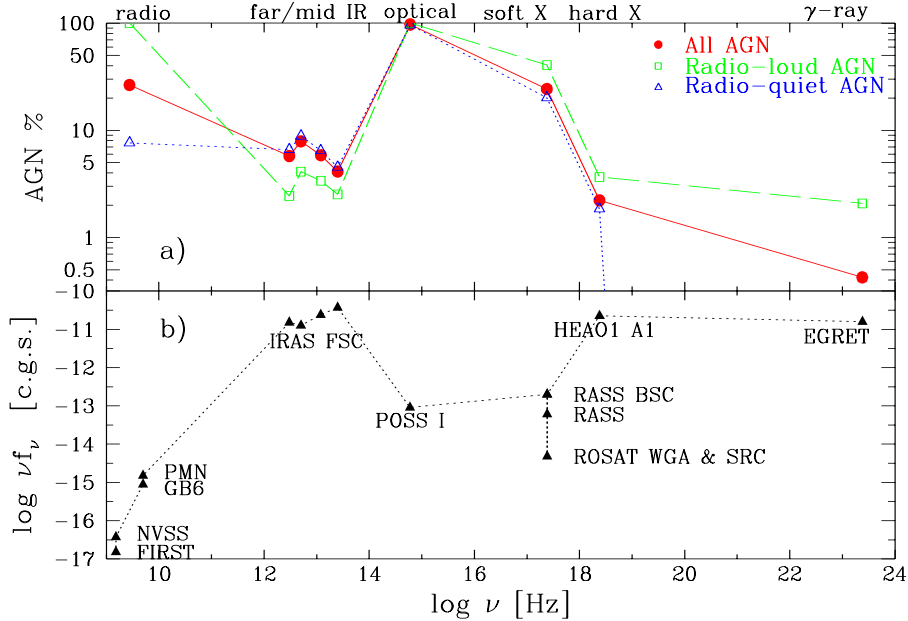,height=8.5cm,width=12.5cm}
\caption{a) (top) The percentage of AGN with detections in a given band
vs. frequency for all objects ({\it solid points}), radio-loud ({\it open
squares}), and radio-quiet ({\it open triangles}) AGN. Sources without radio
data are included in the latter class. b) (bottom) The sensitivity of
the deepest all-sky (or very large area) surveys as a function of frequency.} 
\end{figure}

Figure 1a, which summarizes more than 30 years of observations, shows the
percentage of AGN with data in a given band as a function of frequency for all
AGN and for the RL and RQ subclasses. It can be seen that the multifrequency
spectrum of AGN is not that well known. We are not doing too badly in the
radio (at a few GHz) and soft X-rays, where about 30\% of AGN are detected,
while basically all AGN have optical data (due to the fact that the eye is
peaked in the optical and we always require an ``optical identification'').
However, we have major ``holes'' in our knowledge in the far/mid-IR, hard X-ray,
and $\gamma$-ray bands. Note also that RL AGN fare better than RQ ones at all
frequencies, apart from the IRAS band. If we restrict ourselves to the
radio/optical/soft X-ray bands, where we have the best coverage, only about
$10\%$ of AGN have data in all three bands. 

The fractions of objects shown in Fig. 1a are the convolution of two effects:
1. the intrinsic spectral shape of AGN; 2. the sensitivity of our detectors in
the various bands. The latter is shown in Fig. 1b, which displays the limiting
sensitivities of the deepest all-sky (or large-area) surveys available at
various frequencies, which were used to construct Fig. 1a (deeper, small-area
surveys exist in some bands). The POSS I limit, as discussed above, is only
indicative as many optical data come from dedicated observations. The main
point here is that the bands where we have less information, i.e. far/mid-IR,
hard X-ray and $\gamma$-ray, are exactly those where our surveys are less
sensitive. It then follows that our poor knowledge of the multifrequency
spectra of AGN is due, more than to their intrinsic faintness in some bands,
to the limitations of our detectors. (The radio band is an exception: radio
surveys are very deep but most AGN are weak radio sources.) 
 
\section{The role of multiwavelength sky surveys in AGN research}
Given the broad-band emission of AGN, it is clear that multiwavelength sky
surveys are extremely important to further our understanding of AGN physics,
at a zero order simply by providing more data for more AGN. 

This is not as trivial as it might sound: AGN are rare, making up only 
about 1\% of all bright galaxies (although low luminosity AGN [e.g., LINERs]
could be relatively more numerous). The latest AGN catalog includes about
12,000 sources, while at this meeting catalogs with tens and even hundreds of
millions of entries have been discussed (e.g., \cite{C97}). AGN are also 
hard to find, especially in the optical. Some of the complete samples, 
which are needed to study AGN evolution, are still quite small, especially the
radio-selected ones, and only less than 300 BL Lacs are known over the whole sky
(\cite{PG95}). The cross-correlation between surveys will take advantage of the
broad-band emission of AGN, enhancing the efficiency of AGN detection,
especially important for rare classes of objects (\cite{N97} and \cite{P97}). 

Specific examples of open questions in AGN research which will benefit from
existing and future sky surveys include the following: 

\begin{itemize}
\item The radio-loud/radio-quiet dichotomy. Despite years of effort, we still
do not know what makes an AGN radio-loud nor why the $f_{\rm r}/f_{\rm B}$ 
distribution is bimodal for optically-selected samples. Deep radio surveys,
like the NVSS and the FIRST, will allow us to study for the first time large
numbers of radio-selected {\it radio-quiet} objects (\cite{G96}), to check if
the $f_{\rm r}/f_{\rm B}$ distribution is still bimodal when the selection is
done in the radio band, and even to study the radio evolution of RQ quasars. 

\item Thermal versus non-thermal emission. While we have strong evidence in 
favour of a dominance of non-thermal (synchrotron and inverse Compton)
emission in radio-loud AGN, the situation is not that clear for radio-quiet 
sources, although the optical/UV might be dominated by thermal emission. 
Multifrequency data for large numbers of AGN are needed to address this 
question on a statistical basis. 

\item AGN Evolution. The driving force behind the strong evolution observed in 
AGN is still not understood. Also, although the form of the evolution seems
to be similar at radio, optical, and X-ray frequencies, this should be checked
against larger and deeper samples. 

\end{itemize}
 
This list is no doubt incomplete but it is only meant to
give a flavor of the possibilities multiwavelength surveys provide us with. I
will now concentrate on a particular problem, which is a text-book example of
how monochromatic surveys can be misleading: the missing Type 2 QSO. 

\section{The mystery of the missing Type 2 QSO}
According to unified schemes, Type 2 objects are Type 1's seen edge-on, i.e.,
with their central parts strongly obscured by dust. Seyfert 2's have been
identified as the Type 2 equivalent of Seyfert 1's, and radio galaxies as the
Type 2 equivalent of radio quasars. We also think that Seyfert 1 galaxies are
simply low-luminosity, low-redshift versions of RQ quasars. There should then be
Type 2, i.e., narrow-lined, RQ quasars: but where are they? 

\begin{figure}
\psfig{file=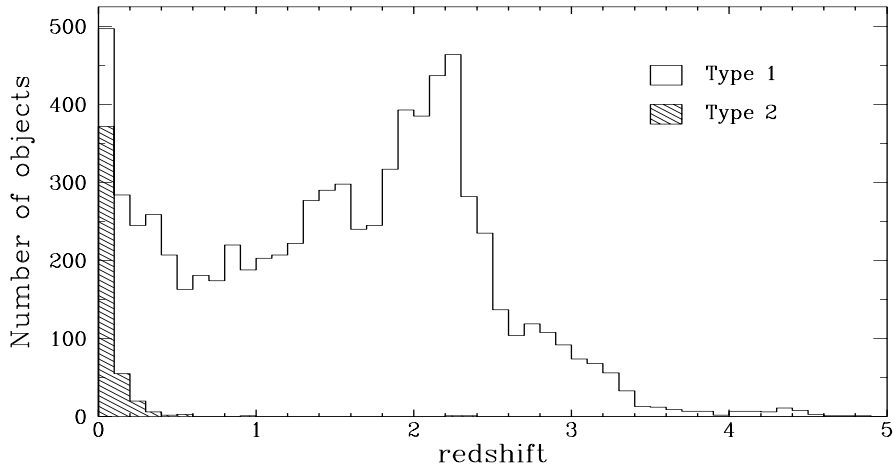,height=6.5cm,width=12.5cm}
\caption{The redshift distribution of the $\sim 8,000$ radio-quiet AGN in the
V\'eron-Cetty \& V\'eron catalog for Type 1 (broad-lined) and Type 2
(narrow-lined) objects. Only 3 Type 2 radio-quiet AGN are known at $z > 0.9$.}
\end{figure}

Figure 2 shows the redshift distribution of all RQ Type 1 and Type 2 AGN in
\cite{VV96}. Most known Type 2's are local, $\sim 80\%$ of them at $z < 0.1$,
and there seem to be no high-$z$ counterparts to the local Seyfert 2's. Note
that for semi-opening angles of the obscuring torus between 30$^{\circ}$ and 
45$^{\circ}$ (derived from various methods), Type 2 AGN should be intrinsically
more numerous than Type 1's by factors between 6.5 and 2.4, that is {\it most RQ
AGN should be of Type 2}. Even for angles as large as 60$^{\circ}$ the two
classes should be equally numerous. The point is that most quasars are still
identified by optical searches, tuned to find AGN with strong non-stellar
continua, i.e. Type 1 quasars. Normally, in fact, candidates are first selected
on the basis of their non-stellar colors, then observed spectroscopically to see
if they show broad lines. The inferred extinction in Type 2 objects can be
larger than 10 magnitudes, reaching even values as high as a few hundred (e.g.,
\cite{GVH94}), so in these objects the AGN continuum is swamped by the host
galaxy light. Type 2 QSO, then, do not even make it to the candidate level, and
even if they did, they would be discarded because they do not show broad lines!
Type 2 radio sources, of course, are selected because of their radio emission
and, in fact, we have examples of radio galaxies at redshifts almost as high as
those of radio quasars. Seyfert 2's, on the other hand, are relatively easy to
find locally where the selection is based on host galaxy properties. For
example, a complete sample of Seyferts can be derived by taking spectra of all
galaxies down to a given magnitude limit. This is of course only feasible for
relatively bright (and therefore local) galaxies (e.g., $B \le 14.5$:
\cite{HB92}). 

How can we find the high-redshift Type 2 QSO? The energy absorbed by the
obscuring material in the optical/UV/soft X-ray bands will have to be
re-emitted in the IR band. Type 2 AGN should then be strong IR emitters. In
fact, optical identification of IRAS sources has produced a large number of
Type 2 AGN, approximately 30\% of known objects, and IRAS selected Type 2's
make up about 50\% of the few sources at $z > 0.1$. Some of these objects have
also, as expected, bolometric luminosities typical of quasars. However, the
redshift distribution in Fig. 2 includes all these IRAS selected Type 2 AGN,
which leads to the next question: Why is IRAS not finding more of them? 
The problem is that the IRAS survey, as illustrated in Fig. 1b, is not very
sensitive: IRAS  has detected only about 8\% of all known AGN at 60$\mu$ (and 
an even smaller fraction in the other bands) and these are mostly local sources,
$\sim 80$\% at $z < 0.1$. IRAS is then sampling only the local universe and the
few high-z Type 2 AGN detected are probably atypically luminous, like IRAS
F10214+4724  ($z = 2.286$), which has recently been shown to be gravitationally
lensed (\cite{BL95}). 

The torus becomes transparent at hard X-ray energies but even the surveys in
this band, made with the HEAO1 observatory, are not very sensitive: only about
2\% of known AGN have hard X-ray data ($\sim 10$ keV) and again most of them
are local sources ($z < 0.1$). 

\section{Towards an unbiased survey}
If our current understanding of AGN is correct, the example I have just given
is not isolated. Most available surveys, in fact, give us a ``biased'' view of
the AGN population, not only because different components
emit at different frequencies, but mainly because most surveys preferentially 
detect some particular classes of AGN. Namely, high-frequency radio
surveys are dominated by objects with beamed radio emission, the so-called
blazars (BL Lacs and core-dominated quasars). Although this is less of a
problem at the faint fluxes reached by the FIRST and NVSS surveys, it can be
shown that even these
surveys will not detect the bulk of the RQ population, so that radio surveys
still preferentially detect RL AGN. (Low-frequency radio surveys are less
biased in terms of selecting beamed objects but still detect mostly RL
sources). The region between near-IR to soft X-ray frequencies is the most
biased of all: here in fact the torus is optically thick and only the nuclear
continuum emission of Type 1 AGN is detected. Radiation from the central parts
manages to escape only at $\lambda \ga 50 - 100\mu$ and $E \ga 10 - 20$ keV,
the precise values depending on the density and geometry of the absorbing column. 
(A
likely distribution of these parameters also implies a distribution in the
energies at which we can see through the torus.) The $\gamma$-ray EGRET sky
survey is also biased, as it has only detected blazars. Far-IR surveys, as
discussed above, are unbiased, but current surveys are not deep enough (also,
they suffer from strong contamination by non-AGN sources, mostly star-forming
regions, both galactic and extragalactic). Finally, hard X-ray surveys are
also unbiased, but still not deep enough. 

What about future far-IR and hard X-ray surveys? As discussed at this meeting
(\cite{B97}), our best bet for a {\it large area}, far-IR survey is the
COBRAS/ SAMBA mission, which will perform all-sky surveys in the $30 -
900$ GHz range. In fact, as regards other IR surveys, they will either be
barely more sensitive than IRAS (e.g., the ISO $200\mu$ serendipitous survey)
or will only sample the mid-IR range (e.g., WIRE) or be limited to small areas
of the sky (e.g., ISO ELAIS \cite{O97}, SIRTF). As regards hard X-rays, there
will be serendipitous 
surveys covering the $2 - 30$ keV (SAX WFC) and $50 - 100$ keV ({\it GRO} BATSE)
ranges but as these instruments were made for other purposes they will not be
much more sensitive than the HEAO1 surveys. On the other hand, the SAX LECS and
MECS will provide a deep serendipitous survey in the $1 - 10$ keV band, while
deep surveys will also be performed in the $0.25 - 12$ keV and $2 - 12$ keV
ranges respectively by XMM (\cite{S97}) and ABRIXAS (\cite{T97}), the former in
serendipitous and slew mode, the latter as a proper all-sky survey. Although
these energy ranges are not optimized to see through the torus in all sources,
the rest-frame frequency increases as (1+z) so that at $z = 1$, for example,
ABRIXAS will sample the $4 - 24$ keV range. These surveys will certainly
represent a major step forward in our quest to obtain an unbiased view of AGN,
independent of orientation. 

In summary, the main conclusions are as follows: 1. both because different
AGN components emit in different bands and because most sky surveys are biased
towards particular AGN classes, no single survey can give us the broad view of
AGN we need to understand them. Therefore, multiwavelength sky surveys are
{\it vital} for AGN research. 2. If our understanding of AGN is correct, we
may be missing between 70\% to 90\% of (radio-quiet) AGN. Deep far-IR and hard
X-ray surveys are needed to see through the absorbing material we think
surrounds the central parts of AGN and identify the large numbers of AGN which
have so far escaped detection. 

{\bf Acknowledgements.} Many thanks to various participants in the IAU 179
Symposium for useful discussions. I also acknowledge helpful comments from two
long-standing collaborators, Meg Urry and Paolo Giommi, the latter being also
responsible for introducing me to the world of database management.

\end{document}